\documentclass[twocolumn,showpacs,superscriptaddress,amsmath,amssymb,letterpaper]{revtex4}
\usepackage[english]{babel}
\usepackage{epsfig}
\usepackage{graphicx}
\usepackage{amsfonts}
\usepackage{eufrak}
\usepackage{fancyheadings}
\usepackage{dcolumn}
\usepackage{bm}

\newcommand{\Bdpipi}{B^{0}_{d}\rightarrow \pi\pi}
\newcommand{\Bdkpi}{B^{0}_{d}\rightarrow K\pi}
\newcommand{\Bskpi}{B^{0}_{s}\rightarrow K\pi}     
\newcommand{\Bskk}{B^{0}_{s}\rightarrow  KK}
\newcommand{\Lbpk}{\Lambda_b^0\rightarrow pK}     
\newcommand{\Lbppi}{\Lambda_b^0\rightarrow p\pi}
\newcommand{\Lbhh}{\Lambda_b^0\rightarrow ph^-}
\newcommand{\Bhh}{B\rightarrow h^+ h'^-}

\newcommand{\lb}{$\Lambda_b^0$ }

\newcommand{\mybf}{}

\begin{document}



\title{Search for $\Lbppi$ and $\Lbpk$ decays in $p\bar{p}$ collisions at $\sqrt{s}~=~1.96\,\rm{TeV}$\\}


\affiliation{Institute of Physics, Academia Sinica, Taipei, Taiwan 11529, Republic of China }
\affiliation{Argonne National Laboratory, Argonne, Illinois 60439 }
\affiliation{Institut de Fisica d'Altes Energies, Universitat Autonoma de Barcelona, E-08193, Bellaterra (Barcelona), Spain }
\affiliation{Istituto Nazionale di Fisica Nucleare, University of Bologna, I-40127 Bologna, Italy }
\affiliation{Brandeis University, Waltham, Massachusetts 02254 }
\affiliation{University of California, Davis, Davis, California 95616 }
\affiliation{University of California, Los Angeles, Los Angeles, California 90024 }
\affiliation{University of California, San Diego, La Jolla, California 92093 }
\affiliation{University of California, Santa Barbara, Santa Barbara, California 93106 }
\affiliation{Instituto de Fisica de Cantabria, CSIC-University of Cantabria, 39005 Santander, Spain }
\affiliation{Carnegie Mellon University, Pittsburgh, PA 15213 }
\affiliation{Enrico Fermi Institute, University of Chicago, Chicago, Illinois 60637 }
\affiliation{Joint Institute for Nuclear Research, RU-141980 Dubna, Russia }
\affiliation{Duke University, Durham, North Carolina 27708 }
\affiliation{Fermi National Accelerator Laboratory, Batavia, Illinois 60510 }
\affiliation{University of Florida, Gainesville, Florida 32611 }
\affiliation{Laboratori Nazionali di Frascati, Istituto Nazionale di Fisica Nucleare, I-00044 Frascati, Italy }
\affiliation{University of Geneva, CH-1211 Geneva 4, Switzerland }
\affiliation{Glasgow University, Glasgow G12 8QQ, United Kingdom }
\affiliation{Harvard University, Cambridge, Massachusetts 02138 }
\affiliation{Division of High Energy Physics, Department of Physics, University of Helsinki and Helsinki Institute of Physics, FIN-00014, Helsinki, Finland }
\affiliation{Hiroshima University, Higashi-Hiroshima 724, Japan }
\affiliation{University of Illinois, Urbana, Illinois 61801 }
\affiliation{The Johns Hopkins University, Baltimore, Maryland 21218 }
\affiliation{Institut f\"ur Experimentelle Kernphysik, Universit\"at Karlsruhe, 76128 Karlsruhe, Germany }
\affiliation{High Energy Accelerator Research Organization (KEK), Tsukuba, Ibaraki 305, Japan }
\affiliation{Center for High Energy Physics: Kyungpook National University, Taegu 702-701; Seoul National University, Seoul 151-742; and SungKyunKwan University, Suwon 440-746; Korea }
\affiliation{Ernest Orlando Lawrence Berkeley National Laboratory, Berkeley, California 94720 }
\affiliation{University of Liverpool, Liverpool L69 7ZE, United Kingdom }
\affiliation{University College London, London WC1E 6BT, United Kingdom }
\affiliation{Massachusetts Institute of Technology, Cambridge, Massachusetts 02139 }
\affiliation{Institute of Particle Physics: McGill University, Montr\'eal, Canada H3A~2T8; and University of Toronto, Toronto, Canada M5S~1A7 }
\affiliation{University of Michigan, Ann Arbor, Michigan 48109 }
\affiliation{Michigan State University, East Lansing, Michigan 48824 }
\affiliation{Institution for Theoretical and Experimental Physics, ITEP, Moscow 117259, Russia }
\affiliation{University of New Mexico, Albuquerque, New Mexico 87131 }
\affiliation{Northwestern University, Evanston, Illinois 60208 }
\affiliation{The Ohio State University, Columbus, Ohio 43210 }
\affiliation{Okayama University, Okayama 700-8530, Japan }
\affiliation{Osaka City University, Osaka 588, Japan }
\affiliation{University of Oxford, Oxford OX1 3RH, United Kingdom }
\affiliation{University of Padova, Istituto Nazionale di Fisica Nucleare, Sezione di Padova-Trento, I-35131 Padova, Italy }
\affiliation{University of Pennsylvania, Philadelphia, Pennsylvania 19104 }
\affiliation{Istituto Nazionale di Fisica Nucleare Pisa, Universities of Pisa, Siena and Scuola Normale Superiore, I-56127 Pisa, Italy }
\affiliation{University of Pittsburgh, Pittsburgh, Pennsylvania 15260 }
\affiliation{Purdue University, West Lafayette, Indiana 47907 }
\affiliation{University of Rochester, Rochester, New York 14627 }
\affiliation{The Rockefeller University, New York, New York 10021 }
\affiliation{Istituto Nazionale di Fisica Nucleare, Sezione di Roma 1, University di Roma ``La Sapienza," I-00185 Roma, Italy }
\affiliation{Rutgers University, Piscataway, New Jersey 08855 }
\affiliation{Texas A\&M University, College Station, Texas 77843 }
\affiliation{Texas Tech University, Lubbock, Texas 79409 }
\affiliation{Istituto Nazionale di Fisica Nucleare, University of Trieste/\ Udine, Italy }
\affiliation{University of Tsukuba, Tsukuba, Ibaraki 305, Japan }
\affiliation{Tufts University, Medford, Massachusetts 02155 }
\affiliation{Waseda University, Tokyo 169, Japan }
\affiliation{Wayne State University, Detroit, Michigan 48201 }
\affiliation{University of Wisconsin, Madison, Wisconsin 53706 }
\affiliation{Yale University, New Haven, Connecticut 06520 }


\author{D.~Acosta}
\affiliation{University of Florida, Gainesville, Florida 32611 }

\author{J.~Adelman}
\affiliation{Enrico Fermi Institute, University of Chicago, Chicago, Illinois 60637 }

\author{T.~Affolder}
\affiliation{University of California, Santa Barbara, Santa Barbara, California 93106 }

\author{T.~Akimoto}
\affiliation{University of Tsukuba, Tsukuba, Ibaraki 305, Japan }

\author{M.G.~Albrow}
\affiliation{Fermi National Accelerator Laboratory, Batavia, Illinois 60510 }

\author{D.~Ambrose}
\affiliation{Fermi National Accelerator Laboratory, Batavia, Illinois 60510 }

\author{S.~Amerio}
\affiliation{University of Padova, Istituto Nazionale di Fisica Nucleare, Sezione di Padova-Trento, I-35131 Padova, Italy }

\author{D.~Amidei}
\affiliation{University of Michigan, Ann Arbor, Michigan 48109 }

\author{A.~Anastassov}
\affiliation{Rutgers University, Piscataway, New Jersey 08855 }

\author{K.~Anikeev}
\affiliation{Fermi National Accelerator Laboratory, Batavia, Illinois 60510 }

% ...

\collaboration{CDF Collaboration}

\preprint{\tt CDF/PUB/BOTTOM/PUBLIC/7443}
\preprint{Draft 2.1}

\begin{abstract}
We  report  on  a   search  for  $\Lambda_b^0\rightarrow  p\pi^-$  and
$\Lambda_b^0\rightarrow  pK^-$   (and  charge  conjugate)   decays  in
$p\bar{p}$     collisions    at     $\sqrt{s}=1.96\,\rm{TeV}$    using
$193\,\rm{pb^{-1}}$ of data collected  by the CDF~II experiment at the
Fermilab Tevatron Collider.  Data were collected using a track trigger
that has been optimized to select
tracks belonging to a secondary vertex that is typical of 
two   body  charmless  decays   of  $b$-flavored   hadrons,  including
$\Lambda_b^0$ baryons.  As  no \lb signal was observed, we set the
upper limits on the branching fraction $\mathcal{B}(\Lbhh)$, where $h$
is  $K$   or  $\pi$,  of  $2.3~\times~10^{-5}$   at  $90\%$~C.L.   and
$2.9~\times~10^{-5}$ at $95\%$~C.L.
\end{abstract}


\maketitle

Charmless, hadronic $b$-meson decays  have been of great interest because
they provide  important information on  the violation of  the combined
symmetry operations  of charge conjugation  (C) and parity (P)  in the
standard               model               of              electroweak
interactions~\cite{generalfleisher,generalrossner}.      The     first
observation  of  charmless  hadronic   $b$-meson  decays  by  the  CLEO
collaboration   in    1993~\cite{first-Cleo},   and   the   subsequent
realization that hadronic penguin diagrams dominate
some   of  these  decays~\cite{penguin-Cleo},   has  since
stimulated      a       substantial      body      of      theoretical
work~\cite{aftercleo1,aftercleo2,aftercleo3}.     In   contrast,   our
present  theoretical and experimental  knowledge of  the corresponding
b~baryon decays is rather limited.
Measurements  of branching fractions and CP  asymmetries for decays
like $\Lbpk$  or $p\pi$  could provide valuable  new insight  into the
hadronic dynamics of $b$-hadron  decays into charmless final states.  In
the standard model, the  CP-violating rate asymmetries in these decays
are expected to be large  compared to the corresponding asymmetries in
$b$-meson decays~\cite{dunietz,BABAR,BELLE}.

The       existence        of       the       \lb        is       well
established~\cite{ABE-M,ABREU-M,BUSKULIC-M,BARATE-L,ABREU-L,ACKERSTAFF-L},
however, no charmless decays have been observed.
We search for \lb decaying to $p~K$ and $p~\pi$.
Theoretical predictions  for their branching  ratios lie in  the range
$(0.9 - 1.2) \times 10^{-6}$  for $p\pi$ decays and $(1.4 - 1.9)\times
10^{-6}$  for $pK$  decays~\cite{Mohanta}.   The current  experimental
upper  limit on the  branching ratios  of these  decay modes  has been
measured  by  the  ALEPH   experiment  and  is  $5\times  10^{-5}$  at
$90\%$~C.L.~\cite{ALEPHmeas}.  The
hadronic  $b$ trigger  {\mybf  of the  upgraded  Collider Detector  at
Fermilab (CDF~II)} selects events  with track pairs originating from a
common  displaced vertex.  A  clean signal  of charmless  hadronic $B$
decays has  been reconstructed using  this trigger~\cite{Bhhref}.  The
same sample
should contain the two-body charmless \lb decays in $pK$ and $p\pi$.

This  search  uses  a $193\pm12\,\rm{pb}^{-1}$~\cite{Luminosity}  data
  sample recorded by the
CDF~II   experiment   at  the   Tevatron   $p\bar{p}$  collider   with
$\sqrt{s}=1.96\,\rm{TeV}$  between February  2002 and  September 2003.
The components of  the CDF~II detector pertinent to  this analysis are
described   briefly  below.   Detailed   descriptions  can   be  found
elsewhere~\cite{pt-mary}.
 Two    silicon   microstrip    detectors    SVX~II~\cite{SVXII}   and
ISL~\cite{ISL}  and   a  cylindrical  drift   chamber  COT~\cite{COT},
immersed in a $1.4  \,\rm{T}$ solenoidal magnetic field, track charged
particles  in the  range  $|\eta|<1.0$~\cite{CDFcoord}.  The  solenoid
covers  $r<150\,\rm{cm}$.  The  SVX~II  provides up  to five  $r-\phi$
position measurements,  each of $\sim  15 \,\rm{\mu m}$  precision, at
radii between $2.5$ and $10.6$ cm.  The ISL provides one axial and one
stereo measurement with
$\sim 20 \,\rm{\mu  m}$ precision, at radii between  $20$ and $28$ cm,
helping to  connect the tracks in the  COT with those in  the SVX, and
improving the tracking efficiency.  The COT has 96 measurement layers,
between $40  \,\rm{cm}$ and $137 \,\rm{cm}$ in  radius, organized into
eight  alternating axial  and  $\pm 2^\circ$  stereo superlayers.   An
additional   silicon   detector,    L00~\cite{L00},   at   radius   of
$1.3\,\rm{cm}$ is present but is not used in this analysis.

The events used
are selected  with a three-level trigger system.   At Level~1, charged
tracks in  the COT  transverse plane are  reconstructed by  a hardware
processor~(XFT)~\cite{XFT}.   The   trigger  requires  two  oppositely
charged  tracks  with  reconstructed  transverse momenta  $p_T  \ge  2
\,\rm{GeV/c}$ and  $p_{T1}+p_{T2}\ge5.5\,\rm{GeV/c}$.  At Level~2, the
Silicon  Vertex Tracker~(SVT)~\cite{SVT1}  associates  SVX~II position
measurements  with XFT  tracks.   The impact  parameter  of the  track
($d_0$) with  respect to the  beamline is measured with  $50\, \rm{\mu
m}$ resolution,  which includes a $\sim 30\,  \rm{\mu m}$ contribution
from transverse  beam size as  measured in SVT.  Requiring  two tracks
with  $100\,{\rm\mu m}\le |d_0|  \le 1.0\,{\rm  mm}$ selects  a sample
enriched in heavy flavor.  The two trigger tracks must have an opening
angle  between $20^\circ$  and $135^\circ$.   The track  pair  is also
required to  be consistent with  {\mybf originating from }  a particle
having a transverse decay length  larger then $200\,\rm{\mu m}$ and an
impact parameter less then $140\,\rm{\mu m}$.
At Level~3,
{\mybf  we fully reconstruct  the event  using the  offline software.}
Candidate trigger tracks  are then selected from this  improved set of
tracks   by  matching  them   in  curvature   and  $\phi$   to  tracks
reconstructed by the Level~2 trigger.  To select candidate events, the
Level~1 and Level~2 selections are  then applied to the set of matched
tracks,  and the  invariant mass,  assuming the  tracks are  pions, is
required to be within $4$ and $7\,\rm{GeV/c^2}$.

We normalize the $\Lbhh$ branching ratio to the branching ratio
$\mathcal{B}\left(\Bdkpi\right)=\left(1.85\pm         0.11\right)\times
10^{-5}$~\cite{PDG2004}.   The  normalization  mode  has  been  chosen
because  its decay topology  is similar  to that  of the  signal.  The
normalization mode is not well
separated  from the  other  $\Bhh$ decays  at  CDF, namely  $\Bdpipi$,
$\Bskpi$ and $\Bskk$.  To obtain  the yield of $\Bdkpi$ we measure the
overall $\Bhh$ yield
 and then fit  the relative fraction $R =  N(\Bdkpi)/N(\Bhh)$ using an
unbinned maximum likelihood  fit~\cite{Bhhref}.  {\mybf The likelihood
function  has   contributions  from   the  signal  ($\Bhh$)   and  the
background.   The  signal likelihood  is  given  by  the six  distinct
$B^0_{s,d}$ decays  modes into  $KK$, $\pi\pi$ and  $K^\pm\pi^\mp$. In
addition  to   $M_{\pi\pi}$,  the  kinematic  variable   used  is  the
charged-signed       momentum        imbalance,       defined       as
$\alpha=(1-\frac{p_1}{p_2})\cdot q_1$, where  the $p_1\,(p_2)$ are the
modulus of the  smaller (larger) momentum of the  tracks, and $q_1$ is
the charge sign of the track assigned to $p_1$.}
%
%

The  relationship between  the number  of events  ($N$)  and branching
ratios ($\mathcal{B}$)  of the signal  and normalizing mode  are given
by:

\begin{equation}
\mathcal{B}\left(\Lbhh\right)=\frac{N\left(\Lbhh\right)}{A}\quad\quad
\label{eq:1}
\end{equation}
and
\begin{equation}
A=\frac{\epsilon_\Lambda}{\epsilon_B}\cdot   \frac{f_\Lambda}{f_d}\cdot
\frac{R\cdot N\left(\Bhh\right)}{\mathcal{B}\left(\Bdkpi \right)}
\label{eq:2}
\end{equation}

where  $\epsilon_\Lambda$ ($\epsilon_B$) is  the total  efficiency for
observing  a \lb ($B^0_d$)  and $f_\Lambda$  ($f_d$) is  the $b$-quark
hadronization  fraction of  the $\Lambda_b^0$  $(B^0_d)$.  We  use the
following   values:  $f_\Lambda   =   0.099\pm  0.017$   and  $f_d   =
0.397\pm0.010$~\cite{PDG2004flfd}.   These  mean  values are  obtained
from measurements at  both LEP (see~\cite{frac-DELPHI,frac-ALEPH}) and
CDF~\cite{CDF-fragment}, using data  samples containing both $b$ baryons
and mesons  and sensitive to $p_T$ of  the \lb down to  10 GeV/c.  The
value of the ratio we use is
$f_\Lambda/f_d  =   0.25\pm0.04$.    We  estimate   the
efficiencies using a detailed
Monte Carlo simulation of the detector and of the trigger
{\mybf  using}  GEANT~\cite{Geant}, to  generate  samples  of \lb  and
$B^0_d$.

A blind  analysis was  performed. The data  in the signal  mass window
were hidden and the analysis selections optimized without knowledge of
their actual impact  on the result.  The background  was calculated by
fitting the  invariant mass spectrum and interpolating  in the blinded
signal region.  Only  after all selection criteria were  fixed and the
systematic  uncertainties estimated was  the signal  region unblinded,
and  the number  of  events  counted and  compared  with the  expected
background.  Potential  biases in the  background estimate, introduced
by the cut optimization procedure,  were avoided by splitting the full
sample into two  statistically independent sub-samples: one consisting
of even event numbers and the other one of odd event numbers. One half
of  the sample  was used  for  the cut  optimization described  below;
the background level measured on the other half
{\mybf  has  been  multiplied   by  two  to}  calculate  the  expected
 background in the search window.

We select  candidate track pairs from  the set of  offline tracks that
match trigger  tracks based on  invariant mass, impact  parameter, and
transverse decay length of the track pair, as well as impact parameter
of each
 track. The exact criteria are optimized as discussed below.  Figure 1
shows the invariant mass distribution after all selection criteria are
applied.  The dotted line indicates the region that was blinded during
the cut optimization.  The solid line indicates the fit region used to
determine the expected background level.

\begin{figure}[hbt]
\begin{center}
\epsfig{file=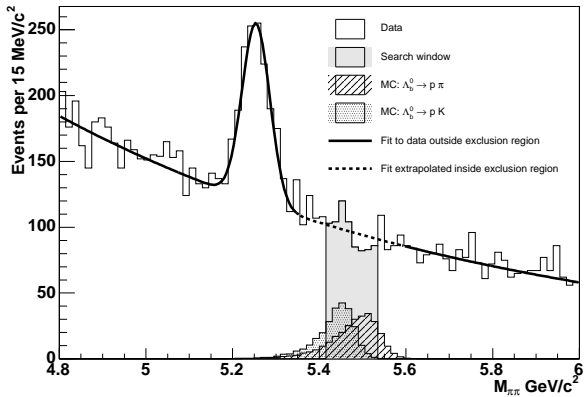,width=\columnwidth}
\end{center}
\small{\caption{\label{fig:plotmassa}    Di   pion    invariant   mass
distribution of all the events including the
search  window. The  function is  the one  from which  we  extract the
number of $\Bhh$ and background events.
The dashed  curve shows the  fitted function in  the part of  the mass
range that was  excluded from the fit.  The scales  of the Monte Carlo
distributions of the  two signal decay modes are  arbitrary.  The peak
in the data is given by the
$\Bhh$
events.}}
\end{figure}

We
assign the  pion mass to all  tracks, resulting in  slight mass shifts
between the various $b$-hadron decays to two-track final states.

{\mybf A  large Monte  Carlo sample including  $\Bhh$ and  $\Lbhh$ was
used to determine  the separation of the two  mass peaks.  For $\Bhh$,
the mean  and rms  were $5258\,\rm{MeV/c^2}$ and  $34\,\rm{MeV/c^2}$ .
For  $\Lbhh$,   the  mean   and  rms  were   $5454\,\rm{MeV/c^2}$  and
$60\,\rm{MeV/c^2}$.     The    separation   is    $196\,\rm{MeV/c^2}$,
sufficiently large}
to make the background from $\Bhh$ negligible within the $\Lambda_b^0$
search window, as can  be easily seen from Figure \ref{fig:plotmassa}.
The   background  in   the   $\Lambda_b^0$  search   window  is   thus
predominantly combinatoric and can be estimated from the sidebands
{\mybf to the right and left of} the search window.

{\mybf The selection criteria, including  the size and position of the
signal region,  were determined  from an optimization  procedure.  The
sideband  regions were  defined to  include those  candidates  with an
invariant  mass between $4.800$  and $5.355\,\rm{GeV/c^2}$  or $5.595$
and $6\,\rm{GeV/c^2}$.   In the optimization} procedure,  we take half
the candidate events in the sideband regions, and maximize a figure of
merit given  by $S/\left(1.5+\sqrt{BKG}\right)$~\cite{punzi} where $S$
and  $BKG$  represent the  number  of  signal  and background  events,
respectively.   The constant  in the  denominator is  chosen  to favor
selections that maximize the
{\mybf sensitivity} reach at $3\sigma$ significance.  This expression
reduces to the usual $S/\sqrt{BKG}$  when the background rate is large
and to $S/1.5$ when the background
is negligible.  Hence observing that the signal is proportional to the
efficiency  ($\epsilon_{\Lambda}$), in  the  optimization we  maximize
$\epsilon_{\Lambda}/\left(1.5+\sqrt{BKG}\right)$ where the efficiency
has been evaluated using the Monte Carlo sample.  We
simultaneously optimize the cuts on
the impact  parameter of  the candidate ($d_\Lambda$),  its transverse
decay length $L_{xy}$  and the minimum impact parameter  of the tracks
($min(d_{01},d_{02})$).  {\mybf  The optimal point has  been found for
$|d_\Lambda|<50\,\rm{\mu    m}$,    $L_{xy}>400\,\rm{\mu    m}$    and
$min(d_{01},d_{02})>180\,\rm{\mu m}$.}   The size and  location of the
mass search window inside the blinded region
{\mybf has  been optimized according to  the same figure  of merit and
spans    the   mass    range    between   $5.415\,\rm{GeV/c^2}$    and
$5.535\,\rm{GeV/c^2}$}.

 
To  estimate the expected  background, we  use the  other half  of the
$\Lambda_b^0$ sideband sample,  and fit it to a sum  of a Gaussian for
the  $\Bhh$  signal,  and  various  combinations  of  exponential  and
polynomial functions for  the combinatoric background.  The systematic
error  in  the  yield  of  $\Bhh$ and  in  the  expected  combinatoric
background  to the  $\Lbhh$ signal  are estimated  from the  spread of
values obtained from different background models.
Table  \ref{tab:system}   summarizes  these  as  well   as  all  other
systematic uncertainties described below.  As the central value we use
the result obtained  with the simplest model consisting  of a Gaussian
plus an exponential distribution.   We arrive at $772\pm31$ events for
the  expected  background  in  the $\Lambda_b^0$  search  window,  and
$726\pm82$  events for  the $\Bhh$  yield. Uncertainties  here include
both the statistical and systematic errors.


\begin{table}[hbt]
\begin{center}
\begin{tabular}{l|l|c} 
\hline \hline  Affected qty. & Source  & Syst. Error  (\%)\\ \hline $B
\rightarrow  h^+  h'^-$   yield  &  Bkg.  shape  &   $5.7$  \\  \hline
Bkg.    estimate   &    Bkg.    shape   &    $3.3$    \\   \hline    &
$(\Lambda_b^0\rightarrow p\pi) /  (\Lambda_b^0\rightarrow pK)$ & $3.5$
\\  &   Window  position  &  $1.2$   \\  &  Window  width   &  $9$  \\
$\epsilon_{\Lambda_b^0}/\epsilon_{B}$ & Lifetime & $3.6$ \\ & proton's
trigger efficiency & $6$ \\ & $p_T(\Lambda_b^0)$ & $17$ \\ & Overall &
{\mybf    $21$}    \\   \hline    \multicolumn{2}{l|}{$\mathcal{B}(B_d
\rightarrow        K\pi)$}       &        $5.9$        \\       \hline
\multicolumn{2}{l|}{$f_{\Lambda}/f_d$} & $16$ \\ \hline \hline
\end{tabular}
\end{center}
\caption{\label{tab:system}  List  of  the relative  systematic  error
contributions to the measurement.}
%
\end{table}

To  calculate the  $\mathcal{B}(\Lbhh)$  we need  not  only the  event
yields  from  the  data  but   also  the  ratio  of  the  efficiencies
$\epsilon_B/\epsilon_\Lambda$  which  we  evaluate using  Monte  Carlo
samples of $\Lbhh$ and $\Bdkpi$.
The  efficiency $\epsilon_\Lambda$  was evaluated  assuming  that both
$\Lbppi$ and $\Lbpk$ contribute with the same weight to the signal.
We  estimate a  systematic uncertainty  of $3.5\%$,  allowing  for all
possible  values for  the ratio  of branching  fractions.   {\mybf The
efficiency ratio is also sensitive  to the lifetime of the $b$ hadron,
because   the  trigger   event   selection  depends   on  the   vertex
displacement.}     In   the    simulation,   lifetime    values   from
PDG~\cite{PDG2004} have been used.
We varied the lifetime  values within the experimental uncertainty and
observe a variation in the  efficiency ratio of $3.6\%$. We quote this
as a systematic error.

%
We  assign   additional  systematic  uncertainties   due  to  possible
discrepancies
between Monte Carlo and data with regard to invariant mass scale, mass
resolution, and specific ionization  in the COT for different particle
species. The  resulting discrepancies in the mass  distribution are of
order  a few $\rm{MeV/c^2}$  and influence  signal efficiency  via the
position and width of the search window. Varying position and width of
the  search window  to reflect  the measured  differences of  data and
Monte Carlo leads to variations  in signal efficiency of 1.2\% and 9\%
respectively.

A  third  source of  systematic  error  is  the variation  of  trigger
efficiency  with  particle  species,  which arises  from  a  different
ionization  energy loss in  the tracking  chamber.  We  evaluated this
effect  by  adjusting  the   efficiency  for  pions  and  kaons  using
corrections  obtained from data.   As protons  and kaons  have similar
ionization in the momentum range  of interest, the efficiency for both
has been corrected in the same way.
After  the correction, the  overall variation  {\mybf of  the relative
efficiency}  of $6\%$  was taken  as  the systematic  error from  this
source.
 
The main contribution to the systematic error comes from the potential
difference in $p_T$ spectra  between $b$ mesons and $\Lambda_b^0$.  As
the $\Lambda_b^0$ $p_T$ spectrum is  not well measured, we use the $b$
hadron spectrum from~\cite{pt-mary} and assume that all hadrons (mesons
and baryons) have the same spectrum.
We  compare  the  efficiency  for  the integrated  spectrum  with  the
efficiencies for  two specific $p_T$ values. As  specific $p_T$ values
we use  the mean  of the  $b$ meson $p_T$  distribution, and  the mean
$p_T$ of  the combinatoric background events below  the search window.
We  assign a 17\%  systematic error  based on  the spread  among these
three efficiency estimates.


The  value  of  the  efficiency  ratio  $\epsilon_B/\epsilon_\Lambda$,
corrected  for  the  trigger  efficiency of  different  particles,  is
$1.77\pm0.37$,   where  the  error   includes  both   statistical  and
systematic errors.  The measured value of the factor $A$ is
$(3.2\pm   1.0)\times   10^6$,   where  statistical   and   systematic
uncertainties   are  included,   in  addition   to   uncertainties  on
$\mathcal{B}(\Bdkpi)$  and  on  the  production fractions  ($f_d$  and
$f_\Lambda$)(Eq. \ref{eq:2}).
The fraction of $\Bdkpi$ ($R$ in eq. \ref{eq:2}) is calculated and the
result
is $0.59\pm0.04$.

{\mybf The  total number of  events in the  signal region of  the mass
spectrum  is $767$,  consistent within  the error  with  the predicted
background, $772\pm31$. Because there is  no excess of signal over the
predicted background, we calculate upper limits on the}
number of signal events and the branching ratio
using a  Bayesian method with uniform prior  distribution. This method
takes  into   account  the   effect  of  statistical   and  systematic
uncertainties.
The  resulting upper  limits on  the number  of signal  events  and on
$\mathcal{B}(\Lbhh)$ are respectively  $75$ and $2.3\times 10^{-5}$ at
$90\%$~C.L and $97$ and $2.9\times  10^{-5}$ at $95\%$~C.L.  This is a
significant  improvement  over   the  previously  published  limit  of
$5\times     10^{-5}$     at     $90\%$~C.L     for     both     decay
modes~\cite{ALEPHmeas}.  Substantially  more  statistics and  improved
background   suppression   is   needed   to   reach   the   level   of
$1-2\times10^{-6}$ as  predicted for the branching  fractions in these
decays in the Standard Model.

{\bf Acknowledgments}

We thank  M. Gronau,  J.Rosner, I.Stewart and  Wei-Shou Hou  for their
comments.  We thank the Fermilab staff and the technical staffs of the
participating institutions  for their vital  contributions.  This work
was supported by  the U.S.  Department of Energy  and National Science
Foundation;  the Italian  Istituto Nazionale  di Fisica  Nucleare; the
Ministry  of Education,  Culture,  Sports, Science  and Technology  of
Japan;  the  Natural  Sciences  and Engineering  Research  Council  of
Canada; the  National Science  Council of the  Republic of  China; the
Swiss  National Science  Foundation; the  A.P.  Sloan  Foundation; the
Bundesministerium  f\"ur Bildung  und Forschung,  Germany;  the Korean
Science and Engineering Foundation and the Korean Research Foundation;
the  Particle Physics  and Astronomy  Research Council  and  the Royal
Society, UK; the Russian Foundation for Basic Research; the Comisi\'on
Interministerial de  Ciencia y Tecnolog\'{\i}a, Spain; in  part by the
European   Community's  Human   Potential  Programme   under  contract
HPRN-CT-2002-00292; and the Academy of Finland.


\bibliographystyle{apsrev}
\bibliography{lbhh}

\begin{thebibliography}{35}
\bibitem[{CDF()}]{CDFcoord}
       \bibinfo{note}{CDF uses a cylindrical coordinate system in which $\phi$ is the
         azimuthal angle, $r$ is the radius from the nominal beamline and $z$ points
         in the proton beam direction and is zero at the center of the detector. The
         transverse plane is the plane perpendicular to the $z$ axis. The
         pseudorapidity $\eta$ is defined as $\eta \equiv -\ln(\tan(\theta/2)$, where
         $\theta$ is the polar angle measured from $z$ axis.}
\end{thebibliography}

\end{document}